\documentclass[conference]{IEEEtran}
\usepackage[utf8]{inputenc}
\usepackage{amssymb}

\usepackage{cite}
\usepackage{amsmath,amssymb,amsfonts}
\usepackage{algorithmic}
\usepackage{graphicx}
\usepackage{textcomp}
\usepackage{xcolor}
\usepackage{url}
\usepackage{pgfplots}
\usetikzlibrary{patterns}
\pgfplotsset{compat=1.18}
\usepackage{tikz}
\usetikzlibrary{shapes,arrows,positioning,fit,backgrounds}

\usepackage{comment}
\usepackage{booktabs,tabularx,makecell,xcolor,pifont,array}

\newcolumntype{L}{>{\raggedright\arraybackslash\hspace{0pt}}X} 
\usepackage{url} 
\usepackage{hyperref}
\usepackage[caption=false,font=footnotesize]{subfig} 

\usepackage{subcaption} 

\usepackage[ruled,vlined,linesnumbered]{algorithm2e}

\usepackage{caption}
\usepackage{graphicx}
\usepackage[caption=false,font=footnotesize]{subfig} 

\usepackage{booktabs,tabularx,ragged2e}
\newcolumntype{Y}{>{\RaggedRight\arraybackslash}X}

\def\BibTeX{{\rm B\kern-.05em{\sc i\kern-.025em b}\kern-.08em
    T\kern-.1667em\lower.7ex\hbox{E}\kern-.125emX}}
\begin{document}




\title{User-Centric Phishing Detection: A RAG and LLM-Based Approach}


\author{
\IEEEauthorblockN{
Abrar Hamed Al Barwani,
Abdelaziz Amara Korba,
Raja Waseem Anwar
}
\IEEEauthorblockA{
German University of Technology in Oman, Sultanate of Oman
}
}

\maketitle

\begin{abstract}
The escalating sophistication of phishing emails necessitates a shift beyond traditional rule-based and conventional machine-learning-based detectors. Although large language models (LLMs) offer strong natural language understanding, using them as standalone classifiers often yields elevated false-positive (FP) rates, which mislabel legitimate emails as phishing and create significant operational burden. This paper presents a personalized phishing detection framework that integrates LLMs with retrieval-augmented generation (RAG). For each message, the system constructs user-specific context by retrieving a compact set of the user’s historical legitimate emails and enriching it with real-time domain and URL reputation from a cyber-threat intelligence platform, then conditions the LLM’s decision on this evidence. We evaluate four open-source LLMs (Llama4-Scout, DeepSeek-R1, Mistral-Saba, and Gemma2) on an email dataset collected from public and institutional sources. Results show high performance; for example, Llama4-Scout attains an F1-score of 0.9703 and achieves a 66.7\% reduction in FPs with RAG. These findings validate that a RAG-based, user-profiling approach is both feasible and effective for building high-precision, low-friction email security systems that adapt to individual communication patterns.
\end{abstract}

\begin{IEEEkeywords}
Phishing Detection, False Positives, Large Language Models (LLMs), Retrieval-Augmented Generation (RAG), Personalized Spam Filter, Email Security
\end{IEEEkeywords}

\section{Introduction}
Email phishing remains one of the most pervasive and financially damaging cyber threats, with organizations facing relentless social engineering attacks that bypass conventional security controls. Traditional detection mechanisms~\cite{salloum2022systematic}, including signature-based filters, blacklists, and heuristic analysis—provide essential but increasingly inadequate protection against evolving attack vectors. These approaches suffer from fundamental limitations: static rule sets unable to adapt to novel tactics, high false positive rates that burden users, and insufficient contextual understanding of sophisticated social engineering techniques.

The emergence of artificial intelligence has introduced transformative capabilities for addressing these challenges~\cite{chinta2025building}. Traditional machine learning approaches have demonstrated promising results in classifying malicious emails through feature extraction and pattern recognition. However, the advent of large language models (LLMs) has fundamentally advanced the landscape of cybersecurity~\cite{diaf2024bartpredict, tellache2025advancing, dif2025towards} and email security. Specifically, LLM-based solutions~\cite{desolda2024, koide2024, lee2025} have shown remarkable capabilities in natural language understanding, contextual analysis, and reasoning about social engineering tactics. These systems leverage sophisticated pattern recognition and semantic analysis to identify subtle phishing indicators that often escape traditional detection methods.

Despite these advancements, a critical research gap persists in leveraging LLMs for personalized phishing detection that minimizes false positives. Current AI-powered email security systems \cite{desolda2024, koide2024, lee2025} predominantly operate as generic classifiers, analyzing emails in isolation without considering individual user context. This approach leads to excessive false alarms when legitimate emails deviate from generic patterns but align with specific user behaviors and communication histories. The absence of user-specific context prevents these systems from distinguishing between universally malicious content and communications that are unusual yet legitimate for particular individuals.

This paper introduces a novel framework that addresses this research gap through Retrieval-Augmented Generation (RAG) enhanced LLMs specifically designed for personalized phishing detection. Our system leverages historical email patterns and real-time threat intelligence to create adaptive, user-centric detection capabilities. The fundamental innovation of our approach lies in the dual-context retrieval mechanism that combines \textit{user-specific historical patterns} with \textit{real-time threat intelligence}. This enables the system to distinguish between genuinely malicious emails and legitimate communications that may appear suspicious to generic classifiers but are normal for specific users. The main contributions of this work are as follows:

\begin{itemize}
    \item Introducing a personalized RAG-enhanced framework that integrates user-specific email history with real-time threat intelligence to create contextual awareness previously absent in LLM-based detection systems. The proposed architecture incorporates semantic similarity search, dynamic context retrieval, and multi-source evidence integration to reduce false positives while maintaining high detection accuracy.

    \item Developing a comprehensive evaluation methodology that systematically assesses four diverse open-source language models across both standalone and RAG-enhanced configurations. This rigorous testing protocol establishes a standardized framework for benchmarking AI-powered security tools.

    \item Demonstrating significant empirical improvements through experimental evidence showing substantial false positive reduction, with Llama4-Scout achieving a 4\% FPR with RAG compared to 12\% without. The analysis reveals consistent performance gains across model sizes and architectures, confirming that personalization benefits scale effectively from 17B to 70B parameter models.
\end{itemize}

The remainder of this paper is organized as follows. Section~\ref{sec:rl} reviews related work. Section~\ref{sec:ps} details the proposed solution. Section~\ref{sec:per} presents the performance evaluation and results. Section~\ref{sec:con} concludes the paper and outlines directions for future work.

\section{Related Works}~\label{sec:rl}

Early approaches to phishing email detection primarily relied on traditional machine learning techniques and rule-based systems. Gupta et al. \cite{gupta2018} comprehensively surveyed existing solutions including DNS-based blacklists, Sender ID, Domain Keys, and machine learning algorithms like k-Nearest Neighbor. These methods, while foundational, suffered from significant limitations including high false positive rates, inability to detect zero-day phishing attacks, and dependence on static patterns that could be easily evolved by attackers. The fundamental limitation of these traditional approaches lies in their inability to understand semantic content and contextual nuances of modern phishing emails, which increasingly employ sophisticated social engineering tactics and personalized content that bypasses signature-based detection.

The emergence of large language models has revolutionized the approach to email security by enabling deep semantic understanding of email content. Several recent studies have demonstrated the effectiveness of LLMs in phishing detection: Desolda et al. \cite{desolda2024} developed APOLLO, a GPT-4o powered tool that classifies emails and generates explanations for users. Their system incorporates URL enrichment through VirusTotal and geographical analysis via BigDataCloud. While achieving high accuracy, the authors noted limitations including prompt quality dependency and the need for evaluation with multiple LLM models beyond GPT-4o.

Koide et al. \cite{koide2024} proposed ChatSpamDetector, a system that processes .eml files and leverages prompt engineering with both normal and simple prompt templates. Their evaluation across GPT-4, GPT-3.5, Llama 2, and Gemini Pro showed GPT-4 achieving 99.70\% accuracy with the normal prompt. However, the study excluded emails without links and suggested that Retrieval-Augmented Generation could further enhance performance. Catherine Lee \cite{lee2025} implemented a comprehensive framework for phishing email detection using multiple open-source LLMs including Llama-3.1-70b, Gemma2-9b, and Mistral-large-latest. The study demonstrated that LLMs could achieve over 80\% accuracy, with Llama-3.1-70b reaching 97.21\%. A key limitation identified was the tendency for "aggressive phishing classification" leading to false positives on imbalanced datasets.

Heiding et al. \cite{heiding2024} investigated the generation of phishing emails using LLMs, comparing GPT-4 generated emails with manually crafted emails using the V-Triad model. Their findings revealed that while LLMs could generate convincing phishing content, human-crafted emails using psychological principles achieved higher click-through rates. This highlights the importance of understanding social engineering tactics in detection systems. Zhang et al. \cite{zhang2025} focused on the practical deployment of LLMs for small and midsize enterprises, finding that smaller models like Llama-3-8b-instruct could provide cost-effective solutions while maintaining robust detection capabilities. Their work challenged the "bigger-is-better" assumption in LLM selection for security applications.

Retrieval-Augmented Generation has emerged as a powerful technique to enhance LLM performance by providing external, up-to-date knowledge. While RAG has been widely adopted in question-answering systems and knowledge-intensive tasks, its application in cybersecurity, particularly phishing detection, remains underexplored. The fundamental advantage of RAG lies in addressing key limitations of standalone LLMs, including hallucination, static knowledge cutoffs, and lack of domain-specific context \cite{lewis2020rag}. By integrating dynamic knowledge retrieval, RAG enables LLMs to make decisions based on current threat intelligence and user-specific historical patterns.

Despite these advancements, current LLM-based phishing detection systems share a critical limitation: they operate as generic classifiers without incorporating user-specific context. This results in high false positive rates when legitimate emails deviate from general patterns but align with individual user behaviors and communication histories. The absence of personalization prevents existing systems from distinguishing between universally malicious content and communications that are unusual yet legitimate for specific users. Table~\ref{tab:related_works} synthesizes recent state-of-the-art contributions in LLM-based phishing detection, highlighting their principal findings and limitations.

Our work addresses this gap by introducing a RAG-enhanced framework that leverages both user-specific email history and real-time threat intelligence, creating a personalized detection system that significantly reduces false positives while maintaining high detection accuracy.

\begin{table*}[t]
\caption{Summary of Key Studies in LLM-based Phishing Detection}
\label{tab:related_works}
\setlength{\tabcolsep}{6pt}
\begin{tabularx}{\textwidth}{@{} l c Y Y @{}}
\toprule
\textbf{Study} & \textbf{Year} & \textbf{Main Findings} & \textbf{Limitations} \\
\midrule
Desolda et al.~\cite{desolda2024} & 2024 & High accuracy with GPT-4o, quality explanations & Prompt dependency, single model evaluation \\
Koide et al.~\cite{koide2024}   & 2024 & 99.70\% accuracy with GPT-4, comprehensive analysis & Excluded emails without links, no RAG integration \\
Lee et al.~\cite{lee2025}     & 2025 & Multi-model evaluation, 97.21\% accuracy with Llama-3.1-70b & Aggressive classification, false positives on imbalanced data \\
Heiding et al.~\cite{heiding2023phishing} & 2024 & Comparative analysis of LLM vs human-crafted phishing & Limited dataset, rapid model evolution \\
Zhang et al.~\cite{zhang2025}   & 2025 & Cost-effective solutions for SMEs, smaller model effectiveness & Limited against sophisticated attacks \\
\bottomrule
\end{tabularx}
\end{table*}

\section{Proposed Solution}~\label{sec:ps}

\begin{figure*}[!ht]
\centering
\includegraphics[width=0.75\linewidth]{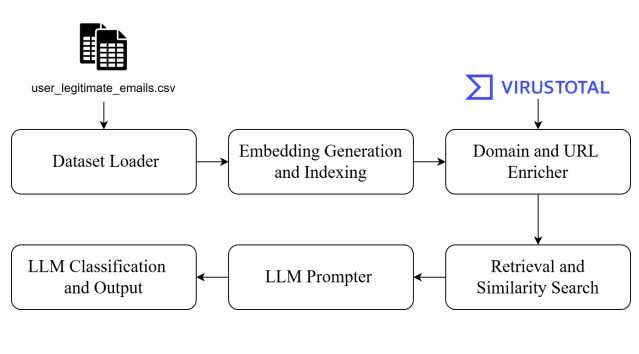}
\caption{Architecture of the RAG-based LLM Phishing Detection System}
\label{fig:system_architecture}
\end{figure*}

Our proposed framework introduces a novel RAG-enhanced architecture for personalized phishing detection that addresses the critical limitation of false positives in AI-based systems. The system operates through six stages to provide context-aware email classification, as illustrated in Figure~\ref{fig:system_architecture}.

\subsection{Data Preprocessing and Feature Extraction Pipeline}
The system initialization commences with a rigorous data preprocessing pipeline that transforms raw email corpora into a structured knowledge base. Let $\mathcal{D}_{\text{raw}} = \{e_1, e_2, \dots, e_n\}$ represent the raw email dataset, where each email $e_i$ comprises heterogeneous features. Our preprocessing pipeline applies a sequence of transformation functions:

\begin{equation}
\mathcal{D}_{\text{clean}} = \Phi_{\text{validate}} \circ \Phi_{\text{normalize}} \circ \Phi_{\text{structure}} \circ \Phi_{\text{decode}}(\mathcal{D}_{\text{raw}})
\end{equation}

where the transformation operators are defined as:

\begin{align}
\Phi_{\text{decode}}(e) &:= \text{Multi-encoding resolution using} \\
&\quad \{\text{UTF-8}, \text{Latin-1}, \text{ISO-8859-1}\}, \\
\Phi_{\text{structure}}(e) &: \text{Feature selection}
\{\text{subject}, \text{sender}, \text{body}\} \subset \mathcal{F} \\
\Phi_{\text{normalize}}(e) &: \text{Unicode normalization and header stripping}, \\
\Phi_{\text{validate}}(e) &:= \mathbf{1}_{\{\text{sender} \in \mathcal{E}_{\text{valid}}\}} \cdot e,
\quad \text{where } \mathcal{E}_{\text{valid}} := \{\, x \mid \texttt{@} \in x \,\}.
\end{align}

The resultant cleaned dataset $\mathcal{D}_{\text{clean}}$ serves as the foundational corpus for personalized context retrieval, ensuring maximal information preservation while eliminating noise and inconsistencies. The pseudocode underlying the data-processing pipeline is illustrated in Algorithm~\ref{alg:preprocessing}.

\begin{algorithm}
\caption{Data Preprocessing Pipeline}
\label{alg:preprocessing}
\begin{algorithmic}[1]
\REQUIRE $\mathcal{D}_{\text{raw}} = \{e_1, e_2, \dots, e_n\}$
\ENSURE $\mathcal{D}_{\text{clean}}$
\STATE $\mathcal{D}_{\text{intermediate}} \gets \emptyset$
\FOR{each $e_i \in \mathcal{D}_{\text{raw}}$}
    \STATE $e' \gets \text{DecodeWithFallback}(e_i,$
\STATE \hspace{\algorithmicindent} $\{\text{UTF-8, Latin-1, ISO-8859-1}\})$
    \STATE $e' \gets \text{ExtractFeatures}(e', \{\text{subject, send er, body}\})$
    \STATE $e' \gets \text{RemoveNonASCII}(\text{StripHeaders}(e'))$
    \IF{$\text{ContainsValidEmail}(e'.\text{sender})$}
        \STATE $\mathcal{D}_{\text{intermediate}} \gets \mathcal{D}_{\text{intermediate}} \cup \{e'\}$
    \ENDIF
\ENDFOR
\STATE $\mathcal{D}_{\text{clean}} \gets \text{RemoveIncomplete}(\mathcal{D}_{\text{intermediate}})$
\RETURN $\mathcal{D}_{\text{clean}}$
\end{algorithmic}
\end{algorithm}

\subsection{Semantic Embedding Generation and Vector Space Construction}
This module projects the textual email content into a continuous vector space where semantic relationships are preserved. Let $f_{\theta}: \mathcal{T} \rightarrow \mathbb{R}^d$ be the embedding function parameterized by $\theta$, where $d=384$ for the \texttt{all-MiniLM-L6-v2}~\cite{allminilm-l6-v2} transformer architecture. For each email $e \in \mathcal{D}_{\text{clean}}$, we compute its semantic embedding:
\begin{equation}
\mathbf{v}_e = f_{\theta}(\text{concat}(e.\text{subject}, e.\text{sender}, e.\text{body}))
\end{equation}

The embeddings are L2-normalized to reside on the unit hypersphere:
\begin{equation}
\hat{\mathbf{v}}_e = \frac{\mathbf{v}_e}{\|\mathbf{v}_e\|_2}
\end{equation}

These normalized embeddings are indexed in a FAISS vector database $\mathcal{V}$, optimized for approximate nearest neighbor search with cosine similarity metric:
\begin{equation}
\text{sim}(\mathbf{q}, \mathbf{v}) = \frac{\mathbf{q} \cdot \mathbf{v}}{\|\mathbf{q}\| \|\mathbf{v}\|} = \hat{\mathbf{q}} \cdot \hat{\mathbf{v}}
\end{equation}

\subsubsection{Threat Intelligence Integration via Multi-Engine Analysis}
The system augments semantic analysis with real-time threat intelligence through formal integration with VirusTotal's API. For an incoming email $e_{\text{query}}$, we extract and analyze:

\begin{align}
\mathcal{D}_{\text{domains}} &= \{\text{extract\_domain}(e_{\text{query}}.\text{sender})\} \\
\mathcal{D}_{\text{urls}} &= \{\text{extract\_urls}(e_{\text{query}}.\text{body})\}
\end{align}

Each element $x \in \mathcal{D}_{\text{domains}} \cup \mathcal{D}_{\text{urls}}$ undergoes multi-engine analysis:
\begin{equation}
\mathcal{R}(x) = \left\{ \text{scan}_{i}(x) \right\}_{i=1}^{75} \quad \text{where } \text{scan}_{i} \in \mathcal{E}_{\text{engines}}
\end{equation}

The threat assessment $\mathcal{T}(e_{\text{query}})$ aggregates results across all analyzable elements:
\begin{equation}
\mathcal{T}(e_{\text{query}}) = \bigcup_{x \in \mathcal{D}_{\text{domains}} \cup \mathcal{D}_{\text{urls}}} \left\{ 
\begin{array}{l}
\text{harm}_x, \text{susp}_x, \text{mal}_x, \\
\text{reput}_x
\end{array} 
\right\}
\end{equation}

\subsubsection{Contextual Retrieval via Semantic Similarity Search}
This module enables personalized phishing detection by retrieving semantically similar emails from the user's historical corpus. The process maps emails to a vector space where semantic relationships are preserved, then performs efficient similarity search.

This approach provides the LLM with concrete examples of legitimate emails that share semantic characteristics with the query, enabling personalized classification decisions based on the user's unique communication patterns rather than generic rules.

The pseudocode algorithm~\ref{alg:retrieval} summaries the contextual retrieval steps.

\begin{algorithm}
\caption{Contextual Retrieval Algorithm}
\label{alg:retrieval}
\begin{algorithmic}[1]
\REQUIRE 
    \STATE $e_{\text{query}}$: Input email to classify
    \STATE $\mathcal{V}$: Vector database of historical email embeddings  
    \STATE $k=5$: Number of similar emails to retrieve
\ENSURE $\mathcal{C}_{\text{context}}$: Top-k most similar historical emails
\STATE
\STATE $\mathbf{q} \gets f_{\theta}(e_{\text{query}}.\text{subject} \oplus e_{\text{query}}.\text{sender} \oplus e_{\text{query}}.\text{body})$
\STATE $\hat{\mathbf{q}} \gets \mathbf{q} / \|\mathbf{q}\|_2$ \COMMENT{Normalize to unit sphere}
\STATE $\mathcal{I} \gets \text{FAISS\_kNN}(\mathcal{V}, \hat{\mathbf{q}}, k)$ \COMMENT{Optimized similarity search}
\STATE $\mathcal{C}_{\text{context}} \gets \{ e_i \in \mathcal{D}_{\text{clean}} : i \in \mathcal{I} \}$
\RETURN $\mathcal{C}_{\text{context}}$
\end{algorithmic}
\end{algorithm}

The semantic similarity is computed using cosine similarity in the normalized embedding space:
\begin{equation}
\text{sim}(e_{\text{query}}, e_i) = \frac{\mathbf{q} \cdot \mathbf{v}_i}{\|\mathbf{q}\| \|\mathbf{v}_i\|}
\end{equation}

\subsection{Structured Prompt Engineering and Reasoning Framework}
The LLM prompter module formalizes the classification task through sophisticated prompt engineering. Let $\mathcal{P}$ represent the prompt construction function:

\begin{equation}
\mathcal{P}(e_{\text{query}}, \mathcal{C}_{\text{context}}, \mathcal{T}(e_{\text{query}})) = \text{concat}(r, c_1, c_2, c_3, o)
\end{equation}

where:
\begin{align}
r &: \text{Role assignment: ``cybersecurity expert specialized} \\ &\text{in phishing detection''} \\
c_1 &: \text{Email content: } e_{\text{query}}.\{\text{subject, sender, body}\} \\
c_2 &: \text{RAG context: } \mathcal{C}_{\text{context}} \\
c_3 &: \text{Threat intelligence: } \mathcal{T}(e_{\text{query}}) \\
o &: \text{Output specification enforcing JSON schema } \mathcal{S}
\end{align}

The output schema $\mathcal{S}$ constrains the LLM response to a structured format:
\begin{equation}
\mathcal{S} = \left\{
\begin{aligned}
&\text{Classification\_decision} \in \{\text{legitimate}, \text{phishing}\} \\
&\text{Phishing\_score} \in [0,10] \subset \mathbb{Z} \\
&\text{Risk} \in \{\text{low}, \text{medium}, \text{high}\} \\
&\text{Social\_engineering\_elements} \subseteq \mathcal{A}_{\text{tactics}} \\
&\text{Recommended\_actions} \subseteq \mathcal{A}_{\text{mitigation}} \\
&\text{Brief\_reason} \in \mathcal{T}_{\text{natural}}
\end{aligned}
\right.
\end{equation}

\subsection{Model Selection and Configuration}
We evaluated four diverse open-source LLMs to ensure the generalizability of our approach across different architectures and scales. Table~\ref{tab:model_specs} summarizes the model specifications.

\begin{table*}
\centering
\caption{Specifications of Selected LLM Models}
\label{tab:model_specs}
\begin{tabular}{lcccc}
\toprule
\textbf{Feature} & \textbf{Llama4-Scout} & \textbf{DeepSeek-R1} & \textbf{Mistral-Saba} & \textbf{Gemma2-9B} \\
\midrule
\textbf{Provider} & Meta & DeepSeek & Mistral & Google \\
\textbf{Base Model} & LLaMA 4 & LLaMA 2 & Mistral & Gemma 2 \\
\textbf{Parameters} & 17B & 70B & 24B & 9B \\
\textbf{Context Window} & 131k & 128k & 32k & 8k \\
\textbf{Architecture} & Sparse MoE & Dense & Dense & Transformer \\
\bottomrule
\end{tabular}
\end{table*}

This formalization ensures the LLM grounds its reasoning in the multi-modal evidence while producing structured, machine-parsable outputs that facilitate automated processing and integration.
\subsection{LLM Classification and Output}
The final module handles LLM inference and output processing:

\begin{itemize}
\item \textbf{API Integration}: We utilize Groq~\cite{groq-lpu-architecture} API for high-speed inference with low latency, crucial for real-time email classification
\item \textbf{Parameter Configuration}: Temperature is set to 0.2 to ensure deterministic, consistent responses while maintaining necessary reasoning capability
\item \textbf{System Message}: A persistent system message reinforces the LLM's role and behavioral guidelines throughout the interaction
\item \textbf{Result Parsing}: The JSON output is automatically parsed and validated, with error handling for malformed responses
\end{itemize}

The module outputs both the classification decision and supporting analysis, providing users with transparent, explainable results.

\section{Experimental Setup and Results} \label{sec:per}
In this section, we first describe the email dataset collection and processing, then detail the implementation of our system, and finally present the experimental results and discuss the main findings.
\subsection{Dataset collection and processing.}
We assembled a balanced corpus of 500 emails (250 legitimate, 250 phishing). Legitimate messages were sampled from consenting users’ personal and institutional mailboxes~\cite{gutech} using IMAP in read-only mode, while phishing messages were sourced from public phishing repositories and internal security feeds. Messages were parsed at the MIME level, converted to UTF-8 (NFC), and reduced to three fields—\textit{subject}, \textit{sender}, and \textit{body}—for model input. HTML content was normalized to plain text (tag stripping and entity unescaping), boilerplate footers and quoted replies were heuristically pruned, and tracking artifacts were removed. We applied light normalization (lowercasing, whitespace compaction) to preserve semantic cues relevant to LLMs. To limit personally identifiable information, the local part of email addresses was anonymized and only domain-level information was retained; attachments were ignored. We de-duplicated near-identical messages via a hash of the normalized body and \textit{subject+sender} pairs, filtered malformed samples, and retained only messages with non-empty bodies. When threat-intelligence enrichment was enabled, URLs and domains were extracted (using robust URL parsing) and summarized into a compact reputation snippet appended to the RAG context. For evaluation, we used stratified splits and prevented leakage by (i) excluding the query message from retrieval, and (ii) building the FAISS index only from the legitimate emails in the training portion for each run. 


\begin{figure}
  \centering
  \subfloat[F1-score: With vs Without RAG\label{fig:f1}]{%
    \includegraphics[width=0.48\textwidth]{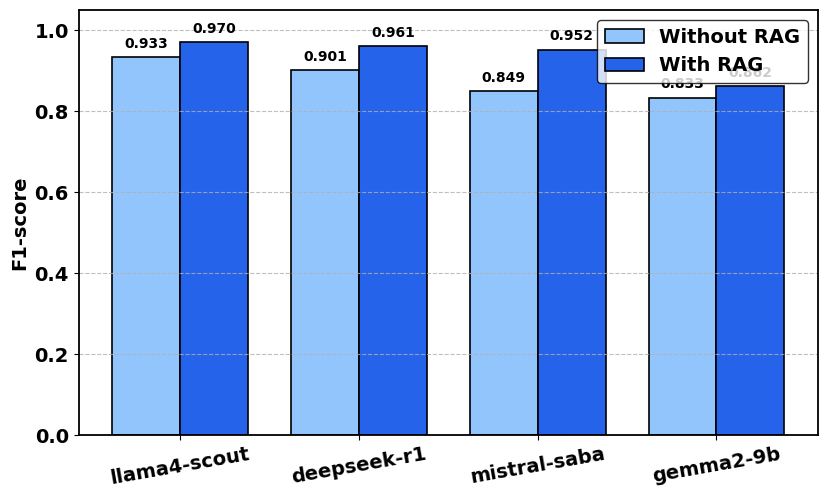}}
    \hfill
    \subfloat[FPR: With vs Without RAG\label{fig:fpr}]{%
    \includegraphics[width=0.48\textwidth]{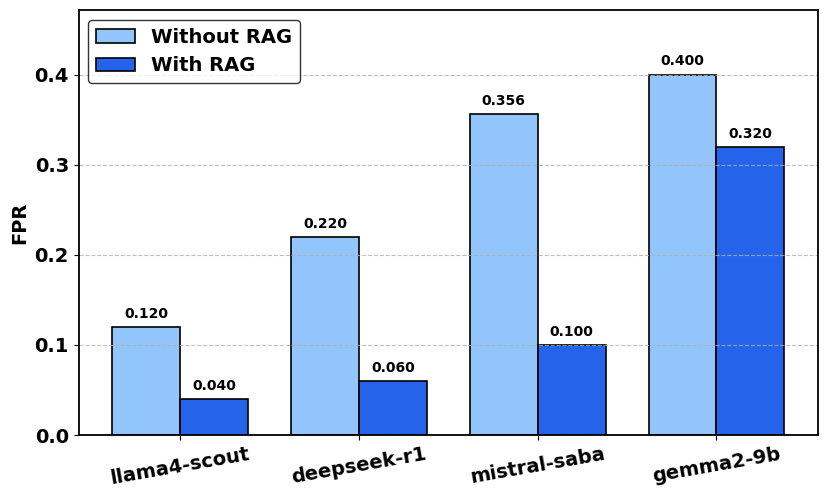}}
  
  \caption{Comparaison des modèles avec et sans RAG.}
  \label{fig:rag_comparison}
\end{figure}
\subsection{Implementation Details}
All models were accessed through GroqCloud~\cite{groqcloud} API to ensure consistent inference speed and reliability, with identical prompt templates and parameter settings (temperature=0.2) across all experiments.
The system was implemented in Python~3.10 and leverages a compact toolchain: LangChain~\cite{langchain} to orchestrate the RAG pipeline and manage embeddings, FAISS~\cite{douze2024faiss} for efficient vector similarity search, \textbf{Sentence-Transformers} for generating text embeddings, Groq~\cite{groq-lpu-architecture} for low-latency LLM inference, and Pandas~\cite{reback2020pandas} for data preprocessing and manipulation. The implementation emphasizes modularity and extensibility, enabling straightforward integration of additional context sources or alternative LLM providers while preserving the core personalization architecture.


\begin{table*}
  \centering
  \caption{Metrics With vs Without RAG Context (N=500)}
  \label{tab:rag_merged}
  \resizebox{\linewidth}{!}{%
  \begin{tabular}{l*{5}{cc}}
    \toprule
    & \multicolumn{2}{c}{\textbf{Accuracy}} 
    & \multicolumn{2}{c}{\textbf{Recall}} 
    & \multicolumn{2}{c}{\textbf{Precision}} 
    & \multicolumn{2}{c}{\textbf{F1-score}} 
    & \multicolumn{2}{c}{\textbf{FPR}} \\
    \cmidrule(lr){2-3}\cmidrule(lr){4-5}\cmidrule(lr){6-7}\cmidrule(lr){8-9}\cmidrule(lr){10-11}
    \textbf{Model} & \textbf{w/o} & \textbf{w/} & \textbf{w/o} & \textbf{w/}
                   & \textbf{w/o} & \textbf{w/} & \textbf{w/o} & \textbf{w/}
                   & \textbf{w/o} & \textbf{w/} \\
    \midrule
    llama4-scout & 0.9300 & 0.9700 & 0.9800 & 0.9800 & 0.8909 & 0.9608 & 0.9333 & 0.9703 & 0.1200 & 0.0400 \\
    deepseek-r1  & 0.8900 & 0.9600 & 1.0000 & 0.9800 & 0.8197 & 0.9423 & 0.9009 & 0.9608 & 0.2200 & 0.0600 \\
    mistral-saba & 0.8220 & 0.9500 & 1.0000 & 1.0000 & 0.7375 & 0.9091 & 0.8489 & 0.9524 & 0.3560 & 0.1000 \\
    gemma2-9b    & 0.8000 & 0.8400 & 1.0000 & 1.0000 & 0.7143 & 0.7576 & 0.8333 & 0.8621 & 0.4000 & 0.3200 \\
    \bottomrule
  \end{tabular}%
  }
\end{table*}

\subsection{Results and Discussion}
As illustrated by the paired bar plots in Fig.~\ref{fig:rag_comparison} (see FPR in Fig.~\ref{fig:fpr} and F1 in Fig.~\ref{fig:f1}), the visual trends align with Table~\ref{tab:rag_merged}: with \emph{RAG}, FPR bars consistently contract while F1 bars expand across models, indicating fewer false alarms with a stronger overall balance. The shift is most conspicuous for \textit{mistral-saba}, whose bars move markedly in the desired direction, whereas \textit{gemma2-9b}—though improved—remains relatively prone to false positives. Consistent with the table, \textbf{llama4-scout with RAG} stands out by pairing the shortest FPR bar with one of the tallest F1 bars, with \textit{deepseek-r1 with RAG} close behind.

As summarized in Table~\ref{tab:rag_merged}, adding \emph{RAG} context markedly improves decision quality across all models: false positives drop, precision rises, and F1 strengthens, while recall was already high without \emph{RAG}. This pattern suggests better decision calibration via contextual disambiguation—fewer spurious triggers with no loss in detection capability. Among the evaluated configurations, \textbf{llama4-scout with RAG} delivers the most compelling performance, combining the best precision–recall balance with the lowest propensity for false alarms. \textbf{deepseek-r1 with RAG} follows closely and is a strong alternative when minimizing false positives is paramount. \textbf{mistral-saba} sees the largest relative benefit from \emph{RAG}—evidence that a model initially “generous” in alerts can be substantially stabilized by context—yet it remains slightly behind the top two. By contrast, \textbf{gemma2-9b}, though improved, remains more prone to false positives and is less suitable where analyst workload must be tightly controlled. In short, \textbf{llama4-scout with RAG} is the recommended choice for superior overall performance and smoother operational use.

\section{Conclusion} \label{sec:con}
We presented a novel, privacy-aware \emph{LLM+RAG} pipeline for phishing email detection that grounds model decisions in user-specific mailbox context and lightweight threat intelligence. Across four heterogeneous open models on a balanced email benchmark, the system consistently reduced false positives while preserving high recall, yielding clear gains in precision and F1-score. These results support the feasibility of retrieval-augmented decisioning for operational email screening and highlight persistent stressors in long, multi-threaded conversations, aggressively obfuscated URLs and domains, and domain shifts that can dilute retrieval quality. Looking ahead, we focus on two priorities: strengthening robustness and generalization via multilingual, institution-scale corpora and targeted ablations of retrieval, prompt, and model choices under domain shift; and improving operational reliability and efficiency through calibrated uncertainty for cost-aware triage and lightweight pipelines (retrieval pruning and distilled small models) to maintain low latency and a modest computational footprint.



\bibliographystyle{IEEEtran}
\bibliography{ref} 

@misc{heiding2023phishing,
  author        = {Heiding, F. and Schneier, B. and Vishwanath, A. and Bernstein, J.},
  title         = {Devising and Detecting Phishing: Large Language Models vs. Smaller Human Models},
  year          = {2023},
  note          = {arXiv:2308.12287},
  url           = {https://arxiv.org/abs/2308.12287}
}

@article{gupta2018,
  title={Defending against phishing attacks: taxonomy of methods, current issues and future directions},
  author={Gupta, B. B. and Arachchilage, N. A. and Psannis, K. E.},
  journal={Telecommunication Systems},
  volume={67},
  pages={247--267},
  year={2018}
}

@article{desolda2024,
  title={APOLLO: A GPT-based tool to detect phishing emails and generate explanations that warn users},
  author={Desolda, G. and Greco, F. and Vigano, L.},
  journal={arXiv preprint arXiv:2410.07997},
  year={2024}
}

@article{koide2024,
  title={Chatspamdetector: Leveraging large language models for effective phishing email detection},
  author={Koide, T. and Fukushi, N. and Nakano, H. and Chiba, D.},
  journal={arXiv preprint arXiv:2402.18093},
  year={2024}
}

@article{lee2025,
  title={Enhancing Phishing Email Identification with Large Language Models},
  author={Lee, Catherine},
  journal={arXiv preprint arXiv:2502.04759},
  year={2025}
}

@article{heiding2024,
  title={Devising and detecting phishing emails using large language models},
  author={Heiding, F. and Schneier, B. and Vishwanath, A. and Bernstein, J. and Park, P. S.},
  journal={IEEE Access},
  year={2024}
}

@article{zhang2025,
  title={Benchmarking and Evaluating Large Language Models in Phishing Detection for Small and Midsize Enterprises: A Comprehensive Analysis},
  author={Zhang, J. and Wu, P. and London, J. and Tenney, D.},
  journal={IEEE Access},
  year={2025}
}

@inproceedings{lewis2020rag,
  title={Retrieval-augmented generation for knowledge-intensive nlp tasks},
  author={Lewis, Patrick and Perez, Ethan and Piktus, Aleksandra and Petroni, Fabio and Karpukhin, Vladimir and Goyal, Naman and K{\"u}ttler, Heinrich and Lewis, Mike and Yih, Wen-tau and Rockt{\"a}schel, Tim and others},
  booktitle={Advances in Neural Information Processing Systems},
  volume={33},
  pages={9459--9474},
  year={2020}
}

@misc{groqcloud,
  author       = {Groq, Inc.},
  title        = {GroqCloud: Low-Latency Inference Platform for Large Language Models},
  year         = {2025},
  url          = {https://groq.com/groqcloud},
  note         = {Accessed: 2025-10-30}
}

@misc{gutech,
  author       = {{German University of Technology in Oman (GUtech)}},
  title        = {German University of Technology in Oman},
  year         = {2025},
  url          = {https://www.gutech.edu.om},
  note         = {Accessed: 2025-10-30}
}

@misc{langchain,
  author       = {LangChain},
  title        = {LangChain: Open-Source Framework for Building LLM Applications},
  year         = {2025},
  url          = {https://www.langchain.com/},
  note         = {Accessed: 2025-10-30}
}

@article{douze2024faiss,
      title={The Faiss library},
      author={Matthijs Douze and Alexandr Guzhva and Chengqi Deng and Jeff Johnson and Gergely Szilvasy and Pierre-Emmanuel Mazaré and Maria Lomeli and Lucas Hosseini and Hervé Jégou},
      year={2024},
      eprint={2401.08281},
      archivePrefix={arXiv},
      primaryClass={cs.LG}
}

@misc{groq-lpu-architecture,
  author = {Groq, Inc.},
  title  = {LPU Architecture: Deterministic, Low-Latency Inference},
  year   = {2025},
  url    = {https://groq.com/lpu-architecture},
  note   = {Accessed: 2025-10-30}
}

@software{reback2020pandas,
  author    = {The pandas development team},
  title     = {pandas-dev/pandas: Pandas},
  month     = feb,
  year      = {2020},
  publisher = {Zenodo},
  version   = {latest},
  doi       = {10.5281/zenodo.3509134},
  url       = {https://doi.org/10.5281/zenodo.3509134}
}

@article{salloum2022systematic,
  title={A systematic literature review on phishing email detection using natural language processing techniques},
  author={Salloum, Said and Gaber, Tarek and Vadera, Sunil and Shaalan, Khaled},
  journal={Ieee Access},
  volume={10},
  pages={65703--65727},
  year={2022},
  publisher={IEEE}
}

@article{chinta2025building,
  title={Building an Intelligent Phishing Email Detection System Using Machine Learning and Feature Engineering},
  author={Chinta, Purna Chandra Rao and Moore, Chethan Sriharsha and Karaka, Laxmana Murthy and Sakuru, Manikanth and Bodepudi, Varun and Maka, Srinivasa Rao},
  journal={European Journal of Applied Science, Engineering and Technology},
  volume={3},
  number={2},
  pages={41--54},
  year={2025}
}

@misc{allminilm-l6-v2,
  author       = {{Sentence-Transformers}},
  title        = {all-MiniLM-L6-v2},
  howpublished = {\url{https://huggingface.co/sentence-transformers/all-MiniLM-L6-v2}},
  year         = {2021},
  note         = {Accessed: 2025-10-31}
}

@inproceedings{diaf2024bartpredict,
  title={Bartpredict: Empowering iot security with llm-driven cyber threat prediction},
  author={Diaf, Alaeddine and Korba, Abdelaziz Amara and Karabadji, Nour Elislem and Ghamri-Doudane, Yacine},
  booktitle={GLOBECOM 2024-2024 IEEE Global Communications Conference},
  pages={1239--1244},
  year={2024},
  organization={IEEE}
}

@article{tellache2025advancing,
  title={Advancing autonomous incident response: Leveraging llms and cyber threat intelligence},
  author={Tellache, Amine and Korba, Abdelaziz Amara and Mokhtari, Amdjed and Moldovan, Horea and Ghamri-Doudane, Yacine},
  journal={arXiv preprint arXiv:2508.10677},
  year={2025}
}

@inproceedings{dif2025towards,
  title={Towards Trustworthy Agentic IoEV: AI Agents for Explainable Cyberthreat Mitigation and State Analytics},
  author={Dif, Meryem Malak and Bouchiha, Mouhamed Amine and Korba, Abdelaziz Amara and Ghamri-Doudane, Yacine},
  booktitle={2025 IEEE 50th Conference on Local Computer Networks (LCN)},
  pages={1--10},
  year={2025},
  organization={IEEE}
}

\end{document}